\documentclass{article}
\usepackage{amsmath}
\usepackage{harvard}

\input{tcilatex}

\begin{document}

\title{Local Causality, Probability and Explanation}
\author{Richard A. Healey \\
The University of Arizona}
\maketitle

\begin{abstract}
In papers published in the 25 years following his famous 1964 proof John
Bell refined and reformulated his views on locality and causality.\ Although
his formulations of local causality were in terms of probability, he had
little to say about that notion. But assumptions about probability are
implicit in his arguments and conclusions. Probability does not conform to
these assumptions when quantum mechanics is applied to account for the
particular correlations\ Bell argues are locally inexplicable. This account
involves no superluminal action and there is even a sense in which it is
local, but it is in tension with the requirement that the direct causes and
effects of events are nearby.
\end{abstract}

\section{Introduction}

I never met John Bell, but his writings have supplied me with a continual
source of new insights as I read and reread them over 40 years. In working
toward a rather different understanding of quantum mechanics he has been
foremost in my mind as a severe but honest critic of such attempts. We all
would love to know what Einstein would have made of Bell's theorem. I
confess the deep regret I feel that Bell cannot respond to this paper is
sometimes assuaged by a sense of relief.

\section{Locality and Local Causality}

In his seminal 1964 paper, John Bell expressed locality as the requirement

\begin{quote}
that the result of a measurement on one system be unaffected by operations
on a distant system with which it has interacted in the past. [2004, p. 14]
\end{quote}

This seems to require that the result of a measurement would have been the
same, no matter what operations had been performed on such a distant system.
But suppose the result of a measurement were the outcome of an
indeterministic process. Then the result of the measurement might have been
different even if exactly the same operations (if any) had been performed on
that distant system. So can no indeterministic theory satisfy the locality
requirement? Bell felt no need to address that awkward question in his 1964
paper, since he took\ the EPR argument to establish that any additional
variables needed to restore locality and causality would have to determine a
unique result of a measurement. Indeterminism was not an option:

\begin{quote}
Since we can predict in advance the result of measuring any chosen component
of $\mathbf{\sigma }_{2}$ [in the Bohm-EPR\ scenario], by previously
measuring the same component of $\mathbf{\sigma }_{1}$, it follows that the
result of any such measurement must actually be predetermined. (\textit{op.
cit.}, p. 15)
\end{quote}

Afterwards he repeatedly stressed that any theory proposed as an attempt to
complete quantum theory while restoring locality and causality need not be 
\textit{assumed} to be deterministic: to recover such perfect
(anti)correlations it would \textit{have} to be deterministic. This argument
warrants closer examination, and I'll come back to it. But in later work
Bell offered formulations of locality conditions tailored to theories that
were not deterministic.

\qquad An initial motivation may have been to facilitate experimental tests
of attempted local, causal completions of quantum mechanics by suitable
measurements of spin or polarization components on pairs of separated
systems represented by entangled quantum states. Inevitable apparatus
imperfections would make it impossible to confirm a quantum prediction of
perfect (anti)correlation for matched components, so experiment alone could
not require a local, causal theory to reproduce them. Einstein himself
thought some theory might come to underlie quantum mechanics much as
statistical mechanics underlies thermodynamics. In each case there would be
circumstances in which the more basic theory (correctly) predicts deviations
from behavior the less basic theory leads one to expect.

\qquad But by 1975 a second motivation had become apparent---the hope that
by revising or reformulating quantum mechanics as a theory of local beables
one might remove ambiguity and arrive at increased precision. It is in this
context that Bell now introduces a requirement of local causality. This
differs in two ways from his earlier locality requirement. It is not a
requirement on the world, but on theories of local beables: and it applies
to theories that are probabilistic, with deterministic theories treated as a
special case in which all probabilities are 0 or 1 (and densities are delta
functions).\footnote{%
While Bell did not make this explicit in 1975, his 1990 paper also notes the
analogy with source-free Maxwellian electromagnetism, and there he does say:
\par
\begin{quote}
The deterministic case is a limit of the probabilistic case, the
probabilities becoming delta functions. [2004, p. 240]
\end{quote}
} Bell ([1975], [1985]) designs his requirement of local causality as a
generalization of a requirement of local determinism met by Maxwell's
electromagnetic theory. In source-free Maxwellian electromagnetism, the
local beables are the values of the electric and magnetic fields at each
point $(\mathbf{x},t)$. This theory is locally deterministic because the
field values in a space-time region are uniquely determined by their values
at an earlier moment in a finite volume of space that fully closes the
backward light cone of that region.

\qquad Local causality arises by generalizing to theories in which the
assignment of values to some beables $\Lambda $ implies, not necessarily a
particular value, but a probability distribution $Pr(A|\Lambda )$, for
another beable $A$. Here is how Bell ([2004, p. 54]) defines it (in my
notation):

\begin{quote}
Let $N$ denote a specification of \textit{all} the beables, of some theory,
belonging to the overlap of the backward light cones of space-like separated
regions 1 and 2. Let $\Lambda $ be a specification of some beables from the
remainder of the backward light cone of 1, and $B$ of some beables in the
region 2. Then in a \textit{locally causal theory} $Pr(A|\Lambda
,N,B)=Pr(A|\Lambda ,N)$ whenever both probabilities are given by the theory.
\end{quote}

If $M$ is a specification of some beables from the backward light cone of 2
but not of 1, then (assuming the joint probability distribution $%
Pr(A,B|\Lambda ,M,N)$ exists)%
\begin{eqnarray}
Pr(A,B|\Lambda ,M,N) &=&Pr(A|\Lambda ,M,N,B).Pr(B|\Lambda ,M,N)  \label{1} \\
&=&Pr(A|\Lambda ,N).Pr(B|\Lambda ,M)  \label{2}
\end{eqnarray}%
where (\ref{1}) follows from the definition of conditional probability, and (%
\ref{2}) follows for any locally causal theory. This means that any theory
of local beables that is locally causal satisfies the condition%
\begin{equation}
Pr(A,B|\Lambda ,M,N)=Pr(A|\Lambda ,N).Pr(B|\Lambda ,M)  \label{3}
\end{equation}

\qquad In his 1990 presentation Bell modified his formulation of local
causality, in part in response to constructive criticisms. He also defended
his revised formulation by appeal to an \textit{Intuitive Principle} of
local causality (IP), namely

\begin{quote}
The direct causes (and effects) of events are near by, and even the indirect
causes (and effects) are no further away than permitted by the velocity of
light. [2004, p. 239]
\end{quote}

Here is Bell's revised formulation of \textit{Local Causality} (LC):

\begin{quote}
A theory will be said to be locally causal if the probabilities attached to
values of local beables in a space-time region 1 are unaltered by
specification of values of local beables in a space-like separated region 2,
when what happens in the backward light cone of 1 is already sufficiently
specified, for example by a full specification of local beables in a
space-time region 3 [a thick "slice" that fully closes the backward light
cone of region 1 wholly outside the backward light cone of 2]. (\textit{op.
cit.}, p. 240).
\end{quote}

Bell [1990] then applies this condition to a schematic experimental scenario
involving a linear polarization measurement on each photon in an entangled
pair in which the polarizer setting $a$ and outcome recording $A$ for one
photon occur in region 1, while those ($b,B$ ) for the other photon occur in
region 2. He derives a condition analogous to (\ref{3}) and uses it to prove
a CHSH inequality whose violation is predicted by quantum mechanics for the
chosen entangled state for certain sets of choices of $a,b$.\footnote{%
Both $\lambda $ and $c$ are assumed confined to region 3 (now symmetrically
extended so as also to close the backward light cone of 2): $c$ stands for
the values of magnitudes characterizing the experimental set-up in terms
admitted by ordinary quantum mechanics, while $\lambda $ specifies the
values of magnitudes introduced by the theory supposed to complete quantum
mechanics. It will not be necessary to mention $c$ in what follows.}%
\begin{equation}
Pr(A,B|a,b,c,\lambda )=Pr(A|a,c,\lambda ).Pr(B|b,c,\lambda ) 
\tag{Factorizability}  \label{Factorizability}
\end{equation}%
Not only did this proof not assume the theory of local beables was
deterministic, even this (Factorizability) condition was not assumed but
derived from the reformulated local causality requirement.

Bell's formulation of local causality (LC) has been carefully analyzed by
Norsen [2011], whose analysis has been further improved by Seevinck and
Uffink [2011]. They have focused in their analyses on what exactly is
involved in a sufficient specification of what happens in the backward light
cone of 1. This specification could fail to be sufficient through failing to
mention local beables in 3 correlated with local beables in 2 through a
joint correlation with local beables in the overlap of the backward light
cones of 1 and 2. A violation of a local causality condition that did not
require such a sufficient specification would pose no threat to the
intuitive principle of local causality: specification of beables in 2 could
alter the probabilities of beables in 1 if \textit{unspecified} beables in 3
were correlated with both through a (factorizable) common cause in the
overlap of the backward light cones of 1 and 2. On the other hand, requiring
a specification of \textit{all} local beables in 3, may render the condition
(LC) inapplicable in attempting to show how theories meeting it predict
correlations different from those successfully predicted by quantum theory.%
\FRAME{fhF}{3.6037in}{2.5374in}{0pt}{}{}{Figure}{\special{language
"Scientific Word";type "GRAPHIC";maintain-aspect-ratio TRUE;display
"USEDEF";valid_file "T";width 3.6037in;height 2.5374in;depth
0pt;original-width 7.28in;original-height 5.1102in;cropleft "0";croptop
"1";cropright "1";cropbottom "0";tempfilename
'NDR1RI05.wmf';tempfile-properties "XPR";}}

To see the problem, consider the set-up for the intended application
depicted in Figure 1. $A$,$B$ describe macroscopic events\footnote{%
I use each of `$A$',`$B$' to denote a random variable with values \{$%
V_{A},H_{A}$\}, \{$V_{B},H_{B}$\} respectively. $e_{A}$ ($\bar{e}_{A}$)
denotes the event in region 1 of Alice's photon being registered as
vertically (horizontally) polarized. $e_{B}$ ($\bar{e}_{B}$) denotes the
event in region 2 of Bob's photon being registered as vertically
(horizontally) polarized.}, each usually referred to as the detection of a
photon linearly polarized either vertically or horizontally relative to an $%
a $- or $b$-axis respectively: $a$,$b$ label events at which an axis is
selected by rotating through an angle $a%
{{}^\circ}%
,b%
{{}^\circ}%
$ respectively from some fixed direction in a plane. The region previously
labeled 3 has been relabeled as 3a, a matching region 3b has been added in
the backward light cone of 2, and `3' now labels the entire continuous
\textquotedblleft stack\textquotedblright\ of space-like hypersurfaces right
across the backward light cones of 1 and 2, shielding off these light cones'
overlap from 1,2 themselves. Note that each of 1,$a$ is space-like separated
from each of 2, $b$.

In some theories, a complete specification of local beables in 3 would
constrain (or even determine) the selection events $a$,$b$. But in the
intended application $a$,$b$ must be treated as free variables in the
following sense: in applying a theory to a scenario of the relevant kind
each of $a$,$b$ is to be specifiable independently in a theoretical model,
and both are taken to be specifiable independently of a specification of
local beables in region 3. Since this may exclude some \textit{complete}
theoretical specifications of beables in region 3 it is best not to require
such completeness. Instead, one should say exactly what it is for a
specification to be sufficient.

Seevinck and Uffink [2011] clarify this notion of sufficiency as a
combination of functional and statistical sufficiency, rendering the label $%
b $ and random variable $B$ (respectively) redundant for predicting $%
Pr_{a,b}(A|B,\lambda )$, the probability a theory specifies for beable $A$
representing the outcome recorded in region 1 given beables $a$,$b$
representing the free choices of what the apparatus settings are in
sub-regions of 1,2 respectively, conditional on outcome $B$ in region 2 and
beable specification $\lambda $ in region 3. This implies%
\begin{equation}
Pr_{a,b}(A|B,\lambda )=Pr_{a}(A|\lambda )  \tag{4a}  \label{4a}
\end{equation}%
Notice that $a,b$ are no longer treated as random variables, as befits their
status as the locus of free choice. It would be unreasonable to require a
theory of local beables to predict the\ probability that the experimenters
make one free choice rather than another: but treating $a,b$ as random
variables (as in Bell's formulation of \ref{Factorizability}) would imply
the existence of probabilities of the form $Pr(a|\lambda )$, $Pr(b|\lambda $%
).

By symmetry, interchanging `1' with `2', `$A$' with `$B$' and `$a$' with `b'
implies%
\begin{equation}
Pr_{a,b}(B|A,\lambda )=Pr_{b}(B|\lambda )  \tag{4b}  \label{4b}
\end{equation}

Seevinck and Uffink [2011] offer equations (\ref{4a}) and (\ref{4b}) as
their mathematically sharp and clean (re)formulation of the condition of
local causality. Together, these equations imply the condition%
\begin{equation}
Pr_{a,b}(A,B|\lambda )=Pr_{a}(A|\lambda )\times \ Pr_{b}(B|\lambda ) 
\tag{Factorizability$_{SU}$}  \label{FactorizabilitySU}
\end{equation}%
used to derive CHSH inequalities. Experimental evidence that these
inequalities are violated by the observed correlations in just the way
quantum theory leads one to expect may then be taken to disconfirm Bell's
intuitive causality principle.

In more detail, Seevinck and Uffink [2011] claim that orthodox quantum
mechanics violates the statistical sufficiency conditions (commonly known as
Outcome Independence, following Shimony)%
\begin{eqnarray}
Pr_{a,b}(A|B,\lambda ) &=&Pr_{a,b}(A|\lambda )  \TCItag{5a}  \label{5a} \\
Pr_{a,b}(B|A,\lambda ) &=&Pr_{a,b}(B|\lambda )  \TCItag{5b}  \label{5b}
\end{eqnarray}

while conforming to the functional sufficiency conditions (commonly known as
Parameter Independence, following Shimony) 
\begin{eqnarray}
Pr_{a,b}(A|\lambda ) &=&Pr_{a}(A|\lambda )  \TCItag{6a}  \label{6a} \\
Pr_{a,b}(B|\lambda ) &=&Pr_{b}(B|\lambda )  \TCItag{6b}  \label{6b}
\end{eqnarray}

Statistical sufficiency is a condition employed by statisticians in
situations where considerations of locality and causality simply don't
arise. But in this application the failure of quantum theory to provide a
specification of beables in region 3 such that the outcome $B$ is always
redundant for determining the probability of outcome $A$ (and similarly with
`$A$', `$B$' interchanged) has clear connections to local causality, as
Seevinck and Uffink's [2011] analysis has shown.

In the light of Seevinck and Uffink's [2011] analysis, perhaps Bell's local
causality condition (LC) should be reformulated as follows:

\begin{quote}
(LC$_{SU}$)\qquad A theory is said to be locally causal$_{SU}$ if it
acknowledges a class $R_{\lambda }$ of beables $\lambda $ in space-time
region 3 whose values may be attached independently of the choice of $a$,$b$
and are then sufficient to render $b$ functionally redundant and $B$
statistically redundant for the task of specifying the probability of $A$ in
region 1.
\end{quote}

The notions of statistical and functional redundancy appealed to here are as
follows:

\begin{quote}
For $\lambda \varepsilon R_{\lambda },$ $\lambda $ renders $B$ statistically
redundant for the task of\newline
specifying the probability of $A$ iff $Pr_{a,b}(A|B,\lambda
)=Pr_{a,b}(A|\lambda )$.

For $\lambda \varepsilon R_{\lambda }$, $\lambda $ renders $b$ functionally
redundant for the task of\newline
specifying the probability of $A$ iff $Pr_{a,b}(A|\lambda )=Pr_{a}(A|\lambda
)$.
\end{quote}

Though admittedly less general than (LC), (LC)$_{SU}$ seems less problematic
but just as well motivated by (IP), as applied to the scenario depicted in
Figure 1. If correlations in violation of the CHSH inequality are locally
inexplicable in so far as no theory of local beables can explain them
consistent with (LC), then they surely also count as locally inexplicable in
so far as no theory of local beables can explain them consistent with (LC)$%
_{SU}$. But Bell himself said we should regard his step from (IP) to (LC)
with the utmost suspicion, and that is what I shall do. My grounds for
suspicion are my belief that quantum mechanics \textit{itself} helps us to
explain the particular correlations violating CHSH inequalities that Bell
[2004, pp. 151-2] claimed to be locally inexplicable without action at a
distance. Moreover, that explanation involves no superluminal action, and
there is even a sense in which it is local.

To assess the status of (LC) (or (LC)$_{SU}$) in quantum mechanics one needs
to say first how it is applied to yield probabilities attached to values of
local beables in a space-time region 1 and then what it would be for these
to be altered by specification of values of local beables in a space-like
separated region 2. This is not a straightforward matter. The Born rule may
be correctly applied to yield more than one chance for the same event in
region 1, and there is more than one way to understand the requirement that
these chances be unaltered by what happens in region 2. As we'll see, the
upshot is that while Born rule probabilities do violate (\ref%
{Factorizability}) (or (\ref{FactorizabilitySU})) here, this counts as a
violation of (LC) (or (LC)$_{SU}$) only if that condition is applied in a
way that is not motivated by (IP)'s prohibition of space-like causal
influences.

\section{Probability and chance}

Bell credited his formulation of local causality (LC) with avoiding the
"rather vague notions" of cause and effect by replacing them with a
condition of probabilistic independence. The connection to (IP)'s motivating
talk of `cause' and `effect' is provided by the thought that a cause alters
(and typically raises) the chance of its effect. But this connection can be
made only by using the general probabilities supplied by a theory to supply
chances of particular events.

By chance I mean the definite, single-case probability of an individual
event such as rain tomorrow in Tucson. As in this example, its chance
depends on \textit{when} the event occurs---afterwards, it is always 0 or 1:
and it may vary up until that time as history unfolds. Chance is important
because of its conceptual connections to belief and action. The chance of $e$
provides an agent's best guide to how strongly to believe that $e$ occurs,
when not in a position to be certain that it does.\footnote{%
I use the \textquotedblleft tenseless present\textquotedblright\ rather than
the more idiomatic future tense here for reasons that will soon become clear.%
} And the comparison between $e$'s chances according as (s)he does or does
not do $D$ are critical in the agent's decision about whether to do $D$.
These connections explain why the chance of an event defaults to 0 or 1 when
the agent is in a position to be certain about it---typically, after it does
or doesn't occur.

Probabilistic theories may be useful guides to the chances of events, but
what they directly yield are not chances but general probabilities of the
form $\Pr_{C}(E)$ for an event of type $E$ relative to reference class $C$.
To apply such a general probability to yield the chance of $e$, you need to
specify the type $E$ of $e$ and also the reference class $C$. A
probabilistic theory may offer alternative specifications when applied to
determine the chance of $e$, in which case it becomes necessary to choose
the appropriate specifications. Actuarial tables may be helpful when
estimating the chance that you will live to be 100, but you differ in all
kinds of ways from every individual whose death figures in those tables.
What you want is the most complete available specification of \textit{your}
situation: this may include much irrelevant information, but it's not
necessary to exclude this since it won't affect the chance anyway. In
Minkowski space-time, the conceptual connection between chance ($Ch$) and
the degree of belief ($Cr$) it prescribes is captured in this version of
David Lewis's Principal Principle that implicitly defines chance:\footnote{%
See Ismael [2008]. I have slightly altered her notation to avoid conflict
with my own. Here `$e$' ambiguously denotes both an event and the
proposition that it occurs. $Cr$ stands for credence: an agent's degree of
belief in a proposition, represented on a scale from 0 to 1 and required to
conform to the standard axioms of probability.}

\begin{quote}
The chance of $e$ at $p$, conditional on any information $I_{p}$ about the
contents of $p$'s past light cone satisfies: $Cr_{p}(e/I_{p})=_{df}Ch_{p}(e)$%
.
\end{quote}

Now consider an agent who accepts quantum theory and wishes to determine the
chance of the event $e_{A}$ that the next photon detected by Alice registers
as vertically polarized ($V_{A}$). Assuming that the state $\left\vert \Phi
^{+}\right\rangle =\frac{1}{\sqrt{2}}\left( \left\vert H\right\rangle
\left\vert H\right\rangle +\left\vert V\right\rangle \left\vert
V\right\rangle \right) $ was prepared and the settings $a,b$ chosen long
before, the agent is also in a position to be certain what these were. The
agent is then in a position to use the Born rule to determine the chance of $%
e_{A}$. But that chance must be relativized not just to a time, but
(relativistically) to a space-time point. So consider the following diagram:%
\FRAME{fhF}{3.4688in}{3.039in}{0pt}{}{}{Figure}{\special{language
"Scientific Word";type "GRAPHIC";maintain-aspect-ratio TRUE;display
"USEDEF";valid_file "T";width 3.4688in;height 3.039in;depth
0pt;original-width 5.834in;original-height 5.1067in;cropleft "0";croptop
"1";cropright "1";cropbottom "0";tempfilename
'NDR1RI06.wmf';tempfile-properties "XPR";}}

As this shows, if the outcome in region $\mathbf{2}$ is of type $V_{B}$ then 
$Ch_{p}(e_{A})=Ch_{p^{\prime }}(e_{A})=Ch_{r}(e_{A})=%
{\frac12}%
$, but $Ch_{q}(e_{A})=\cos ^{2}\angle ab$. These are the chances that follow
from application of the Born rule to state $\left\vert \Phi
^{+}\right\rangle $, given settings $a,b$. In each case the event $e_{A}$\
of the next photon detected in $\mathbf{1}$'s registering as vertically
polarized has been specified as of\ type $V_{A}$, and the specification of
the reference class at least includes the state and settings. Specifically,%
\begin{equation*}
Ch_{p}(e_{A})=Pr_{a,b}^{\Phi ^{+}}(V_{A})=||\hat{P}^{A}(V)\Phi ^{+}||^{2}=%
{\frac12}%
.
\end{equation*}%
\begin{equation*}
Ch_{q}(e_{A})=Pr_{a,b}^{\Phi ^{+}}(V_{A}|V_{B})\equiv \frac{Pr_{a,b}^{\Phi
^{+}}(V_{A},V_{B})}{Pr_{b}^{\Phi ^{+}}(V_{B})}=\frac{%
{\frac12}%
\cos ^{2}\angle ab}{%
{\frac12}%
}=\cos ^{2}\angle ab
\end{equation*}%
Note that the reference class used in calculating $Ch_{q}(e_{A})$ is
narrower: it is further restricted by specification of the outcome as of
type $V_{B}$ in region $\mathbf{2}$.

Any agent who accepts quantum theory and is (momentarily) located at
space-time point $x$ should match credence in $e_{A}$ to $Ch_{x}(e_{A})$
because it is precisely the role of chance to reflect the epistemic bearing
of all information accessible at $x$ on facts not so accessible, and to
accept quantum theory is to treat it as an\ expert when assessing the
chances. This is so whether or not an agent is \textit{actually} located at $%
x$---fortunately, since it is obviously a gross idealization to locate the
epistemic deliberations of a physically situated agent at a space-time
point! A hypothetical agent located at $q$ in the forward light cone of
region $\mathbf{2}$ (but not $\mathbf{1}$) has access to the additional
information that the outcome in $\mathbf{2}$ is of type $V_{B}$: so the
reference class used to infer the chance of $e_{A}$ at $q$ from the Born
rule should include that information. That is why $Ch_{q}(e_{A})$ is
determined by the conditional Born probability $Pr_{a,b}^{\Phi
^{+}}(V_{A}|V_{B})$ but $Ch_{p}(e_{A})$ is determined by the unconditional
Born probability $Pr_{a,b}^{\Phi ^{+}}(V_{A})$.

In the special case that the settings $a,b$ coincide (the polarizers are
perfectly aligned) application of the Born rule yields the chances depicted
in Figure 3.\FRAME{fhF}{3.512in}{3.039in}{0pt}{}{}{Figure}{\special{language
"Scientific Word";type "GRAPHIC";maintain-aspect-ratio TRUE;display
"USEDEF";valid_file "T";width 3.512in;height 3.039in;depth
0pt;original-width 5.8963in;original-height 5.0955in;cropleft "0";croptop
"1";cropright "1";cropbottom "0";tempfilename
'NDR1RI07.wmf';tempfile-properties "XPR";}}\newline
Bell ([2004, pp. 240-41]) took this example as a simple demonstration that
ordinary quantum mechanics is not locally causal, crediting the argument of
EPR [1935]. He begins

\begin{quote}
Each of the counters considered separately has on each repetition of the
experiment a 50\% chance of saying `yes'.
\end{quote}

Each of the chances $Ch_{p}(e_{A}),Ch_{p^{\prime }}(e_{B})$ is 
$\frac12$%
, as Bell says: but $Ch_{q}(e_{A})=1$. After noting that quantum theory here
requires a perfect correlation between the outcomes in $\mathbf{1,2}$, he
continues

\begin{quote}
"So specification of the result on one side permits a 100\% confident
prediction of the previously totally uncertain result on the other side. Now
in ordinary quantum mechanics there just \textit{is} nothing but the
wavefunction for calculating probabilities. There is then no question of
making the result on one side redundant on the other by more fully
specifying events in some space-time region $\mathbf{3}$. We have a
violation of local causality."
\end{quote}

\qquad It is true that (\ref{Factorizability}) and (\ref{FactorizabilitySU})
fail here, since $Pr_{a,a}^{\Phi ^{+}}(V_{A}|V_{B})=1$, $Pr_{a,a}^{\Phi
^{+}}(H_{A}|V_{B})=0$, while $Pr_{a,a}^{\Phi ^{+}}(A)=Pr_{a,a}^{\Phi
^{+}}(B)=%
{\frac12}%
$. But does that constitute a violation of local causality? (LC) and (LC$%
_{SU}$) are both conditions straightforwardly applicable to a theory whose
general probabilities yield a \textit{unique} chance for each possible
outcome in $\mathbf{1}$ prior to its occurrence. In the case of quantum
theory, however, the general Born rule probabilities yield \textit{multiple}
chances for each possible outcome in $\mathbf{1}$, each at the same time\
(in the laboratory frame): $Ch_{p}(e_{A})=%
{\frac12}%
$, but $Ch_{q}(e_{A})=1$ (assuming the outcome in $\mathbf{2}$ is of type $%
V_{B}$). When (LC) speaks of \textquotedblleft \textit{the} probabilities
attached to [$e_{A},\bar{e}_{A}$] in a space-time region $\mathbf{1}$ being
unaltered by specification of [$V_{B}$] in a space-like separated region $%
\mathbf{2}$\textquotedblright\ (my italics), which probabilities are these?

\qquad Since the connection to (IP)'s motivating talk of `cause' and
`effect' is provided by the thought that a cause alters the chance of its
effect, (LC) is motivated only if applied to the \textit{chances} of $e_{A},%
\bar{e}_{A}$ in region $\mathbf{1}$. But $Ch_{p}(e_{A})$ is \textit{not}
altered by specification of $V_{B}$ in space-like separated region $\mathbf{2%
}$: its value depends only on what happens in the backward light cone of $%
\mathbf{1}$, in conformity\ to its role in prescribing $Cr_{p}(e_{A})$.$\ $%
Of course $Ch_{q}(e_{A})$ does depend on the outcome in $\mathbf{2}$. If it
did not, it could not fulfill its constitutive role of prescribing $%
Cr_{q}(e_{A}/I_{q})$ no matter what information $I_{q}$ provides about the
contents of $q$'s past light cone. It follows that $Ch_{q}(e_{A})$ is not
altered but \textit{specified} by specification of the result in $\mathbf{2}$%
.

\qquad Only for a hypothetical agent whose world-line has entered the future
light cone of $\mathbf{2}$ at $q$ is it true that specification of the
result in $\mathbf{2}$\ permits a 100\% confident prediction of the
previously totally uncertain result on the other side. A hypothetical agent
at $p$ is not in\ a position to make a 100\% confident prediction: for such
an agent the result in $\mathbf{1}$ remains totally uncertain: what happens
in $\mathbf{2}$ makes no difference to what (s)he should believe, since
region $\mathbf{2}$ is outside the backward light cone of $p$. That is why
it is $Ch_{p}(e_{A})$, not $Ch_{q}(e_{A})$, that says what is certain at $p$%
. Newtonian absolute time fostered the illusion of the occurrence of future
events becoming certain for everyone at the same time---when they occur if
not sooner. Relativity requires certainty, like chance, to be relativized to
space-time points---idealized locations of hypothetical knowers.

\qquad So does ordinary quantum mechanics violate local causality? If
\textquotedblleft the probabilities\textquotedblright\ (LC) speaks of are $%
Pr_{a,b}^{\Phi ^{+}}(V_{A}),$ $Pr_{a,b}^{\Phi ^{+}}(H_{A})$, and the
condition that these be unaltered is understood to be that $Pr_{a,b}^{\Phi
^{+}}(V_{A})=Pr_{a,a}^{\Phi ^{+}}(V_{A}|B),$ $Pr_{a,b}^{\Phi
^{+}}(H_{A})=Pr_{a,a}^{\Phi ^{+}}(H_{A}|B)$, then ordinary quantum mechanics
violates (LC). But if this is all (LC) means, then it is not motivated by
(IP) and its violation does not imply that the quantum world is non-local in
that there are superluminal causal relations between distant events. For
(LC) to be motivated by causal considerations such as (IP),
\textquotedblleft the probabilities\textquotedblright\ (LC) speaks of must
be understood to be chances, including $Ch_{p}(e_{A})$ and $Ch_{q}(e_{A})$.
But neither of these would be altered by the specification of the outcome [$%
V_{B}$] in a space-like separated region, so local causality would then not
be violated. Although it remains unclear exactly how (LC) (or (LC$_{SU}$))
is supposed to be applied to quantum mechanics, one way of applying it is
unmotivated by (IP), while if it is applied in another way quantum mechanics
does \textit{not} violate this local causality condition.

\section{Chance and Causation}

Suppose in the EPR-Bohm scenario that the outcome in region $\mathbf{2}$ had
been of type $H_{B}$ instead of $V_{B}$: then $Ch_{q}(e_{A})$ would have
been 0 instead of 1. Suppose that the polarization axis for the measurement
in region $\mathbf{2}$ had been rotated through $60%
{{}^\circ}%
$: then $Ch_{q}(e_{A})$ would have been 
$\frac14$
or 
$\frac34$%
, depending on the outcome in region $\mathbf{2}$. Or suppose that no
polarization measurement had been performed in region $\mathbf{2}$: then $%
Ch_{q}(e_{A})$ would have been 
$\frac12$%
. This shows that $Ch_{q}(e_{A})$ depends counterfactually on the
polarization measurement in $\mathbf{2}$ and also on its outcome. Another
way to understand talk of \textquotedblleft alteration\textquotedblright\ of
\textquotedblleft the\textquotedblright\ probability of an event of type $%
V_{A}$ is as the difference between the actual value of $Ch_{q}(e_{A})$ and
what its value would have been had the polarization measurement in $\mathbf{2%
}$ or its outcome been different. Don't such counterfactual
\textquotedblleft alterations\textquotedblright\ in $Ch_{q}(e_{A})$\ amount
to \textit{causal} dependence between space-like separated events, in
violation of (IP)? There are several reasons why they do not.

The first reason is that while $Ch_{q}(e_{A})$ would be different in each of
these counterfactual scenarios, in none of them would $Ch_{p}(e_{A})$ differ
from 
$\frac12$%
, so the "local" chance of $e_{A}$ is insensitive to all such counterfactual
variations in what happens in $\mathbf{2}$. If one wishes to infer causal
from counterfactual dependence of \textquotedblleft the\textquotedblright\
chance of a result in $\mathbf{1}$ on what happens in $\mathbf{2}$, then
only one of two relevant candidates for \textquotedblleft
the\textquotedblright\ chance displays such counterfactual dependence. For
those who think of chance as itself a kind of indeterministic cause---a
localized physical propensity whose actualization may produce an effect---$%
Ch_{p}(e_{A})$ seems better qualified for\ the title of \textquotedblleft
the\textquotedblright\ chance of $e_{A}$\ than $Ch_{q}(e_{A})$.

The role of chance in decision provides the second reason. Just as the
chance of $e$ tells you everything you need to\ know to figure out how
strongly to believe $e$, the causal dependence of $e$ on $d$ tells you
everything you need to know about $e$ and $d$ when deciding whether to do $d$
(assuming you are not indifferent about $e$). As Huw Price [2012] put it,
\textquotedblleft causal dependence should be regarded as an analyst-expert
about the conditional credences required by an evidential decision
maker\textquotedblright .

Consider the situation of a hypothetical agent Bob at $p^{\prime }$ deciding
whether to act by affecting what happens in $\mathbf{2}$ to try to get
outcome $e_{A}$ in $\mathbf{1}$. Bob can choose not to measure anything, or
he can choose to measure polarization with respect to any axis $b$. If he
were to measure nothing, $Ch_{q}(e_{A})$ would be 
$\frac12$%
. If he were to measure polarization with respect to the same axis as Alice,
then $Ch_{q}(e_{A})$ would be either 0 or 1, with an equal chance (at his
momentary location $p^{\prime }$) of either outcome. Since he\ can neither
know nor affect which of these chances it will be, he must base his decision
on his best estimate of $Ch_{q}(e_{A})$ in accordance with Ismael's [2008]
Ignorance Principle:

\begin{quote}
\textquotedblleft Where you're not sure about the chances, form a mixture of
the chances assigned by different theories of chance with weights determined
by your relative confidence in those theories.\textquotedblright
\end{quote}

Following this principle, Bob should assign $Ch_{q}(e_{A})$ the estimated
value $%
{\frac12}%
.0+%
{\frac12}%
.1=%
{\frac12}%
$, and base his decision on that. Since measuring polarization with respect
to the same axis as Alice would not raise his estimated chance of securing
outcome $e_{A}$ in $\mathbf{1}$, he should eliminate this option \textit{%
whether or not he could execute it}. His estimated value of $Ch_{q}(e_{A})$
were he to measure polarization with respect to an axis rotated $60%
{{}^\circ}%
$ from Alice's is also 
$\frac12$
($%
{\frac12}%
.%
{\frac14}%
+%
{\frac12}%
.%
{\frac34}%
=%
{\frac12}%
$). Similarly for any other angle. This essentially recapitulates part of
the content of the no-signalling theorems, going back to Eberhard [1978].
Bell [2004, pp. 237-8] shows why manipulation of external fields at $%
p^{\prime }$ or in $\mathbf{2}$ would also fail to alter Bob's estimated
value of $Ch_{q}(e_{A})$.

But what if Bob had simply arranged for the measurement in $\mathbf{2}$ to
have had the different \textit{outcome} $\bar{e}_{B}$? Then $Ch_{q}(e_{A})$
would have been 0 instead of 1. No-one who accepts quantum mechanics can
countenance this counterfactual scenario. The Born rule implies that $%
Pr_{b}^{\Phi ^{+}}(H_{B})=%
{\frac12}%
$, and anyone who accepts quantum mechanics accepts the implication that $%
Ch_{p^{\prime }}(\bar{e}_{B})=%
{\frac12}%
$. So anyone who accepts quantum mechanics will have credence $Cr_{p^{\prime
}}(\bar{e}_{B}/I_{p^{\prime }})=%
{\frac12}%
$ no matter what he takes to happen in the backward light cone of $p^{\prime
}$ (as specified by $I_{p^{\prime }}$).\footnote{%
A unitary evolution $\Phi ^{+}\Rightarrow \Xi ^{+}$\ corresponding to a 
\textit{local} interaction there would still yield $Pr_{b}^{\Xi ^{+}}(H_{B})=%
{\frac12}%
$.} If he accepts quantum mechanics, Bob will conclude that there is nothing
it makes sense to contemplate doing to alter his estimate of $Ch_{p^{\prime
}}(\bar{e}_{B})$, and so there \ is no conceivable counterfactual scenario
in which one in Bob's position arranges for the measurement in $\mathbf{2}$
to have had the different outcome $\bar{e}_{B}$. In general, there is causal
dependence between events in $\mathbf{1}$ and $\mathbf{2}$ only if it makes
sense to speak of an intervention in one of these regions that would affect
a hypothetical agent's estimated chance of what happens in the other. Anyone
who accepts quantum mechanics should deny that makes sense.

Perhaps the most basic reason why counterfactual dependencies between
happenings in region $\mathbf{2}$ and the chance(s) of $e_{A}$ are no sign
of causal dependence is that chances are not beables, and are incapable of
entering into causal relations. That Bell thought they behaved like beables
is suggested by the [1975] paper in which he introduced local causality as a
natural generalization of local determinism:

\begin{quote}
\textquotedblleft In Maxwell's theory, the fields in any space-time region $%
\mathbf{1}$ are determined by those in any space region $V$, at some time $t$%
, which fully closes the backward light cone of $\mathbf{1}$. Because the
region $V$ is limited, localized, we will say the theory exhibits \textit{%
local determinism}. We would like to form some no[ta]tion of \textit{local
causality} in theories which are not deterministic, in which the
correlations prescribed by the theory, for the beables, are
weaker.\textquotedblright\ [2004, p. 53]
\end{quote}

It seems Bell thought the chances prescribed by a theory that is not
deterministic were analogous to the beables of Maxwell's electromagnetism,
so that while local determinism (locally) specified the local[ized] beables
(e.g. fields), local causality should (locally) specify the local[ized] 
\textit{chances} of beables, where those chances (like local beables) are
themselves localized physical magnitudes.

Others have joined Bell in this view of chances as localized physical
magnitudes. But quantum mechanics teaches us that chances are \textit{not}
localized physical propensities whose actualization may produce an effect.
Maudlin says what he means by calling probabilities objective:

\begin{quote}
\textquotedblleft ...there could be probabilities that arise from
fundamental physics, probabilities that attach to actual or possible events
in virtue solely of their physical description and independent of the
existence of cognizers. These are what I mean by \textit{objective
probabilities}.\textquotedblright\ (Beisbart and Hartmann eds., [2011, p.
294])
\end{quote}

Although quantum chances do attach to actual or possible events, they are
not objective in this sense. As we saw, the chance of outcome $e_{A}$ does
not attach to it in virtue solely of its physical description: the \textit{%
chances} of $e_{A}$ attach also in virtue of its space-time relations to
different space-time locations. Each such location offers the epistemic
perspective of a situated agent, even in a world with no such agents. The
existence of these chances is independent of the existence of cognizers. But
it is only because we are not merely cognizers but physically situated
agents that we have needed to develop a concept of chance tailored to our
needs as informationally deprived agents. Quantum chance admirably meets
those needs: an omniscient God could describe and understand the physical
world without it.

While they are neither physical entities nor physical magnitudes, quantum
chances are objective in a different sense. They supply an objective
prescription for the credences of an agent in any physical situation. Anyone
who accepts quantum mechanics is committed to following that prescription.

\section{A view of quantum mechanics}

As I see it [2012], it is not the function of quantum states, observables,
probabilities or the Schr\"{o}dinger equation to represent or describe the
condition or behavior of a physical system to which they are associated.
These elements function in other ways when a quantum model is applied in
predicting or explaining physical phenomena such as non-localized
correlations. Assignment of a quantum state may be viewed as merely the
first step in a procedure that licenses a user of quantum mechanics to
express claims about physical systems in descriptive language and then
warrants that user in adopting appropriate epistemic attitudes toward some
of these claims. The language in which such claims are expressed is not the
language of quantum states or operators, and the claims are not about
probabilities or measurement results: they are about the values of physical
magnitudes, and I'll refer to them as \textit{magnitude claims}. Magnitude
claims were made by physicists and others before the development of quantum
mechanics and continue to be made, some in the same terms, others in terms
newly introduced as part of some scientific advance. But even though quantum
mechanics represents an enormous scientific advance, claims about quantum
states, operators and probability distributions are not magnitude claims.

\qquad The quantum state has two roles. One is in the algorithm provided by
the Born Rule for assigning probabilities to significant claims of the form $%
M_{\Delta }(s)$ : The value of $M$ on $s$ lies in $\Delta $, where $M$ is a
physical magnitude, $s$ is a physical system and $\Delta $ is a Borel set of
real numbers. In what follows, I will call a descriptive claim of the form $%
M_{\Delta }(s)$ a\textit{\ canonical magnitude claim}. For two such claims
the formal algorithm may be stated as follows:%
\begin{equation}
\Pr (M_{\Delta }(s),N_{\Gamma }(s))=Tr(\rho \hat{P}^{M}[\Delta ].\hat{P}%
^{N}[\Gamma ])  \tag{Born Rule}  \label{Born}
\end{equation}

Here $\rho $ represents a quantum state as a density operator on a Hilbert
space $\mathcal{H}_{s}$ and $\hat{P}^{M}[\Delta ]$ is the value for $\Delta $
of the projection-valued measure defined by the unique self-adjoint operator
on $\mathcal{H}_{s}$ corresponding to $M$.

\qquad But the significance of a claim like $M_{\Delta }(s)$ varies with the
circumstances to which it relates. Accordingly, a quantum state plays a
second role by modulating the content of $M_{\Delta }(s)$ or any other
magnitude claim by modifying its inferential relations to other claims.
Because I believe the nature of this modulation of content renders
inappropriate the metaphor of magnitudes corresponding to elements of
reality, I recommend against thinking of magnitudes that figure in canonical
or other magnitude claims as beables, even though many such magnitude claims
are true. But if one insists on calling magnitudes that figure in magnitude
claims beables, these magnitudes are not beables introduced by quantum
mechanics---they are at most beables recognized in its applications.%
\footnote{%
Compare Bell [2004, p. 55].}

\qquad The quantum state is not a beable in this view. Indeed, since none of
the distinctively quantum elements of a quantum model qualifies as a beable
introduced by the theory, quantum mechanics has no beables of its own.
Viewed this way, a quantum state does not describe or represent some new
element of physical reality.\footnote{%
Compare Bell [2004, p. 53]: "...this does not bother us if we do not grant
beable status to the wave-function."} But nor is it the quantum state's role
to describe or represent the epistemic state of any actual agent. A quantum
state assignment is objectively true (or false): in that deflationary sense
a quantum state is objectively real. But its function is not to say what the
world is like but to help an agent applying quantum mechanics to predict and
explain what happens in it. It is physical conditions in the world that make
a quantum state assignment true (or false). True quantum state assignments
are backed by true magnitude claims, though some of these are typically
about physical systems other than that to which the state is assigned.

\qquad Any application of quantum mechanics involves claims describing a
physical situation. While it is considered appropriate to make claims about
where individual particles are detected contributing to the interference
pattern in a contemporary interference experiment, claims about through
which slit each particle went are frequently alleged to be \textquotedblleft
meaningless\textquotedblright . In its second role the quantum state offers
guidance on the inferential powers, and hence the content, of canonical
magnitude claims.

\qquad The key idea here is that even assuming unitary evolution of a joint
quantum state of system and environment, delocalization of system state
coherence into the environment will typically render descriptive claims
about experimental outcomes and the condition of apparatus and other
macroscopic objects appropriate by endowing these claims with enough content
to license an agent to adopt epistemic attitudes toward them, and in
particular to apply the Born Rule. But an application of quantum mechanics
to determine whether this is so will not require referring to any system as
\textquotedblleft macroscopic\textquotedblright , as an \textquotedblleft
apparatus\textquotedblright\ or as an \textquotedblleft
environment\textquotedblright . All that counts is how a quantum state of a
super-system evolves in a model, given a Hamiltonian associated with an
interaction between the system of interest and the rest of that super-system.

\qquad It is important to note that since the formulation of the Born Rule
now involves no explicit or implicit reference to \textquotedblleft
measurement\textquotedblright , Bell's ([2004, pp. 213-31]) strictures
against the presence of the term `measurement' in a precise formulation of
quantum mechanics are met. None of the other proscribed terms `classical',
`macroscopic', `irreversible', or `information' appears in its stead.

\qquad Since an agent's assignment of a quantum state does not serve to
represent a system's properties, her reassignment of a \textquotedblleft
collapsed\textquotedblright\ state on gaining new information represents no
change in that system's properties. That is why collapse is not a physical
process, in this view of quantum mechanics. Nor does the Schr\"{o}dinger
equation express a fundamental physical law: to assign a quantum state to a
system is not to represent its dynamical properties. A formulation of
quantum mechanics has no need to include a statement distinguishing the
circumstances in which physical processes of Schr\"{o}dinger evolution and
\textquotedblleft collapse\textquotedblright\ occur. An agent can use
quantum mechanics to track changes of the dynamical properties of a system
by noting what magnitude claims are significant and true of it at various
times. But quantum mechanics itself does not imply any such claim, even when
an agent would be correct to assign a system a quantum state, appropriately
apply the Born Rule, and conclude that the claim has probability 1.

\qquad Quantum states are relational on this interpretation. When agents
(actually or merely hypothetically) occupy relevantly different physical
situations they should assign different quantum states to one and the same
system, even though these different quantum state assignments are equally
correct. The primary function of Born probabilities is to offer a physically
situated agent authoritative advice on how to apportion degrees of belief
concerning contentful canonical magnitude claims that the agent is not
currently in a position to check. That is why the Born rule should be
applied by differently situated agents to assign different chances to a
single canonical magnitude claim $M_{\Delta }(s)$ about a system $s$ in a
given situation. These different chance assignments will then be equally
objective and equally correct.

\qquad The physical situation of a (hypothetical or actual) agent will
change with (local) time. The agent may come to be in a position to check
the truth-values of previously inaccessible magnitude claims, some of which
may be taken truly to describe outcomes of measurements. If a quantum state
is to continue to provide the agent with good guidance concerning still
inaccessible magnitude claims, it must be updated to reflect these newly
accessible truths. The required alteration in the quantum state is not a
physical process involving the system, in conflict with Schr\"{o}dinger
evolution. What has changed is just the physical relation of the agent to
events whose occurrence is described by true magnitude claims. This is not
represented by a discontinuous change in the quantum state of some model: it
corresponds to adoption of a \textit{new} quantum model that incorporates
additional information, newly accessible to the user of quantum mechanics.

\qquad The preceding paragraphs contained a lot of talk of agents. To
forestall misunderstandings, I emphasize that quantum mechanics is not about
agents or their states of knowledge or belief: A precise formulation of
quantum mechanics will not speak of such things in its models any more than
it will speak of agents' measuring, observing or preparing activities. If
quantum mechanics is about anything it is about the quantum systems, states,
observables and probability measures that figure in its models. Quantum
mechanics, like all scientific theories, was developed by (human) agents for
the use of agents (not necessarily human: while insisting that any agent be
physically situated, I direct further inquiry on the constitution of agents
to cognitive scientists). Trivially, only an agent can apply a theory for
whatever purpose. So any account of a predictive, explanatory or other
application of quantum mechanics naturally involves talk of agents.\pagebreak

\section{How to use quantum mechanics to explain non-localized correlations}

In his [1981] Bell argued that

\begin{quote}
certain particular correlations, realizable according to quantum mechanics,
are locally inexplicable. They cannot be explained, that is to say, without
action at a distance. [2004, pp. 151-2]
\end{quote}

The particular correlations to which Bell refers arise, for example, in the
EPR-Bohm scenario in which pairs of spin 
$\frac12$
"particles" are prepared in a singlet spin state, then at widely separated
locations each element of a pair is passed through a Stern-Gerlach magnet
and detected either in the upper or in the lower part of a screen. By
calling them realizable rather than realized he acknowledged the
experimental difficulties associated with actually producing statistics
supporting them in the laboratory (or elsewhere). Enormous improvements in
experimental technique since 1981 have overcome most of the difficulties
associated with performing a \textquotedblleft
loophole-free\textquotedblright\ test of CHSH or other so-called Bell
inequalities and at the same time provided very strong statistical evidence
for quantum mechanical predictions in analogous experiments. Since the
improvements have been most dramatic for experiments involving polarization
measurements on entangled photons, it is appropriate to refer back to the
experimental scenario discussed by Bell in 1990 ([2004, pp. 232-248]).

\qquad Suppose photon pairs are prepared at a central source in the
entangled polarization state $\Phi ^{+}=1/\sqrt{2}(|HH\rangle +|VV\rangle )$%
, and the photons in a pair are both subsequently detected in coincidence at
two widely separated locations after each has passed through a polarizing
beam splitter (PBS) with axis set at $a,b$ respectively. If a photon is
detected with polarization parallel to this axis, a macroscopic record
signifies \textquotedblleft yes\textquotedblright : if it is detected with
polarization perpendicular to this axis, the record signifies
\textquotedblleft no\textquotedblright . Let the record \textquotedblleft
yes\textquotedblright\ at one location be the event of a magnitude $A$
taking on value $+1$, \textquotedblleft no\textquotedblright\ the event of $%
A $ taking on value $-1$, and similarly for $B$ at the other location. Let $%
a $ be a locally generated signal that quickly sets the axis of the PBS on
the $A $ side to an angle $a%
{{}^\circ}%
$ from some standard direction, and similarly for $b$ on the $B$ side.
Assume this is done so that each of $a$ and $A$'s taking on a value is
space-like separated from each of $b$ and $B$'s taking on a value. In this
scenario quantum mechanics predicts that, for $a%
{{}^\circ}%
=0%
{{}^\circ}%
,a^{\prime }%
{{}^\circ}%
=45%
{{}^\circ}%
,b%
{{}^\circ}%
=22%
{\frac12}%
{{}^\circ}%
,b^{\prime }%
{{}^\circ}%
=-22%
{\frac12}%
{{}^\circ}%
$%
\begin{equation}
E(a,b)+E(a,b^{\prime })+E(a^{\prime },b)-E(a^{\prime },b^{\prime })=2\sqrt{2}
\tag{7}  \label{CHSH violation}
\end{equation}%
where, for example, $E(a,b)\equiv
\Pr_{a,b}(+1,+1)+\Pr_{a,b}(-1,-1)-\Pr_{a,b}(+1,-1)-\Pr_{a,b}(-1,+1)$. This
is in violation of the CHSH inequality%
\begin{equation}
|E(a,b)+E(a,b^{\prime })+E(a^{\prime },b)-E(a^{\prime },b^{\prime })|\leq 2 
\tag{CHSH}
\end{equation}%
that follows from (\ref{FactorizabilitySU}). Bell claims these correlations
are realizable according to quantum mechanics but that they cannot be
explained without action at a distance. While it is generally acknowledged
that quantum mechanics successfully predicts Bell's particular correlations,
demonstrating this will illustrate the present view of quantum mechanics. To
decide whether it also explains them we need to ask what more is required of
an explanation.

\qquad What we take to be a satisfactory explanation has changed during the
development of physics, and we may confidently expect such change to
continue. One who accepts quantum mechanics is able to offer a novel kind of
explanation. Nevertheless, explanations of phenomena using quantum mechanics
may be seen to meet two very general conditions met by many, if not all,
good explanations in physics.

(i) They show that the phenomenon to be explained was to be expected, and

(ii) they say what it depends on.\newline
Quantum mechanics enables us to give explanations meeting both conditions.

\qquad Meeting the first condition is straightforward. Anyone accepting
quantum mechanics can use the \ref{Born} applied to state $\Phi ^{+}$ to
calculate joint probabilities such as $\Pr_{a,b}(A,B)$ and go on to derive (%
\ref{CHSH violation}). So for anyone who accepts quantum mechanics,
violation of the CHSH inequalities is to be expected. But it is worth
showing in more detail how quantum mechanics can be applied to derive (\ref%
{CHSH violation}) because this will help to exhibit the relational nature of
quantum states and probabilities while making it clear that a precise
formulation of quantum mechanics need not use the word `measurement' or any
other term on Bell's list of proscribed words [2004, p. 215].\FRAME{fhF}{%
3.5916in}{3.039in}{0pt}{}{}{Figure}{\special{language "Scientific Word";type
"GRAPHIC";maintain-aspect-ratio TRUE;display "USEDEF";valid_file "T";width
3.5916in;height 3.039in;depth 0pt;original-width 5.6611in;original-height
4.7841in;cropleft "0";croptop "1";cropright "1";cropbottom "0";tempfilename
'NDR1RI08.wmf';tempfile-properties "XPR";}}

\qquad Figure 4 is a space-time diagram depicting space-like separated
polarization measurements by Alice and Bob in regions 1,2 respectively on a
photon pair. Time is represented in the laboratory frame. At $t_{1}$ each
takes the polarization state of the $L-R$ photon pair to be $\Phi ^{+}=1/%
\sqrt{2}(|HH\rangle +|VV\rangle )$. What justifies this quantum state
assignment is their knowledge of the conditions under which the photon pair
was produced---perhaps by parametric down-conversion of laser light by
passage through a non-linear crystal. Such knowledge depends on information
about the physical systems involved in producing the pair. This state
assignment is backed by significant magnitude claims about such systems---if
anything counts as a claim about beables recognized by quantum mechanics,
these do. Then Alice measures polarization of photon $L$ along axis $a%
{{}^\circ}%
$, Bob measures polarization of photon $R$ along axis $b%
{{}^\circ}%
$. Decoherence at the photon detectors licenses both of them to treat the
Born rule measure corresponding to state assignment $\Phi ^{+}$\ as a
probability distribution over significant canonical magnitude claims about
the values of $A,B$.

\qquad At $t_{1}$ Alice and Bob should both assign state $\Phi ^{+}$ and
apply the \ref{Born}\ to calculate the joint probability $\Pr_{a,b}^{\Phi
^{+}}(A,B)$ assigned to claims about the values of magnitudes $A,B$---claims
that we may, but need not, choose to describe as records of polarization
measurements along both $a%
{{}^\circ}%
,b%
{{}^\circ}%
$ axes---and hence the (well-defined) conditional probability $%
\Pr_{a,b}^{\Phi ^{+}}(A,B)/\Pr_{a,b}^{\Phi ^{+}}(B)=|\langle A|B\rangle
|^{2} $. Each will then expect the observed non-localized correlations
between the outcomes of polarization measurements in regions 1,2 when the
detectors are set along the $a%
{{}^\circ}%
,b%
{{}^\circ}%
$ axes. They will expect analogous correlations as these axes are varied,
and so they will expect (\ref{CHSH violation})\ (violating the CHSH
inequality) in such a scenario.

\qquad At $t_{2}$, after recording polarization $V_{B}$ for $R$, Bob should
assign pure state $|V_{B}\rangle $ to $L$ and use the Born Rule to calculate
the probabilities $\Pr_{a}^{|V_{B}>}(A)=|\langle A|V_{B}\rangle |^{2}$ for
Alice to record polarization of $L$ with respect to the $a%
{{}^\circ}%
$-axis. At $t_{2}$, Alice should assign state $\hat{\rho}=%
{\frac12}%
\hat{1}$ to $L$, and use the Born Rule to calculate probability $%
\Pr_{a}^{\rho }(A)=%
{\frac12}%
$ that she will record either polarization of $L$ with respect to the $a%
{{}^\circ}%
$-axis. In this way each forms expectations as to the outcome of Alice's
measurement on the best information available to him or her at $t_{2}$.
Alice's statistics of her outcomes in many repetitions of the experiment are
just what her quantum state $%
{\frac12}%
\hat{1}$ for $L$ led her to expect, thereby helping to explain her results.
Bob's statistics for Alice's outcomes (in many repetitions in which his
outcome is $V_{B}$) are just what his quantum state $|V_{B}\rangle $ for $L$
led him to expect, thereby helping him to explain Alice's results.

\qquad There is no question as to which, if either, of the quantum states $%
|V_{B}\rangle $, $%
{\frac12}%
\hat{1}$ was the \textit{real} state of Alice's photon at $t_{2}$. Neither
of the different probabilities $|\langle A|V_{B}\rangle |^{2}$ or $%
{\frac12}%
$ represents a unique physical propensity at $t_{2}$ of Alice's
outcome---even though neither its chance at $p$ nor its chance at $q$ is
subjective. This discussion applies independent of the time-order in Alice's
frame of the regions 1, 2 in Figure 1: had she been moving away from Bob
fast enough, she would have represented 1 as earlier than 2.

\qquad It is widely acknowledged that one cannot explain a phenomenon merely
by showing that it was to be expected in the circumstances. To repeat a
hackneyed counterexample, the falling barometer does not explain the coming
storm even though it gives one reason to expect a storm in the
circumstances. As a joint effect of a common cause, a symptom does not
explain its other independent effects. But a system's being in a quantum
state at a time is not a symptom of causes specified by the true magnitude
claims that back it since it is not an event distinct from the\ conditions
those claims describe. Each event figuring in Bell's particular correlations
is truly described by a canonical magnitude claim. We may choose to describe
some, but not all, such events as an outcome of a quantum measurement on a
system: the probabilities of many of those events depend counterfactually on
the particular entangled state assigned at $t_{1}$---if that state had been
different, so would these probabilities. But this dependence is not causal.
In quantum mechanics, neither states nor probabilities are the sorts of
things that can bear causal relations: in Bell's terminology, they are not
beables.

\qquad When relativized to the physical situation of an actual or
hypothetical agent, a quantum state assignment is objectively true or
false---which depends on the state of the world. More specifically, a
quantum state assignment is made true by the true magnitude claims that back
it. One true magnitude claim backing the assignment of $|V_{B}\rangle $ to $%
L $ at $q$ reports the outcome of Bob's polarization measurement in region 2
of Figure 1: but there are others, since this would not have been the
correct assignment had the correct state assignment at $p^{\prime }$ been $%
|H_{A}\rangle |V_{B}\rangle $. We also need to ask for the backing of the
entangled state $\Phi ^{+}$.

\qquad There are many ways of preparing state $\Phi ^{+}$, and this might
also be the right state to assign to some naturally occurring photon pairs
that needed no preparation. In each case there is a characterization in
terms of some set of true magnitude claims describing the systems and events
involved: these back the state assignment $\Phi ^{+}$. It may be difficult
or even impossible to give this characterization in a particular case, but
that is just an epistemic problem which need not be solved even by
experimenters skilled in preparing or otherwise assigning this state. $\Phi
^{+}$ will be correctly assigned at $p^{\prime }$ only if some set of true
magnitude claims backing that assignment is accessible from $p^{\prime }$:
events making them true must lie in the backward light-cone of $p^{\prime }$.

\qquad A quantum state counterfactually depends on the true magnitude claims
that back it in somewhat the same way that a dispositional property depends
on its categorical basis. The state $\Phi ^{+}$ may be backed by alternative
sets of true magnitude claims just as a person may owe his immunity to
smallpox to any of a variety of categorical properties. If Walt owes his
smallpox immunity to antibodies, his possession of antibodies does not cause
his immunity: it is what his immunity consists in. No more is the state $%
\Phi ^{+}$ caused by its backing magnitude claims: a statement assigning
state $\Phi ^{+}$ is true only if backed by some true magnitude claims of
the right kind. A quantum state is counterfactually dependent on whatever
magnitude claims back it because backing is a kind of determination or
constitution relation, not because it is a causal relation.

\qquad In this view, a quantum state causally depends neither on the
physical situation of the (hypothetical or actual) agent assigning it nor on
any of its backing magnitude claims. The correct state $|V_{B}\rangle $ to
be assigned to $L$ at $q$ is not causally dependent on anything about Bob's
physical situation even if he happens to be located at $q$: it is not
causally dependent on the outcome of Bob's polarization measurement in
region 2: and it is not causally dependent on how Bob sets his polarizer in
region 2. But a quantum state assignment is not just a function of the
subjective epistemic state of any agent: If Bob or anyone else were to
assign a state other than $|V_{B}\rangle $ to $L$ at $q$ he or she would be
making a mistake.

\qquad The quantum derivation of (\ref{CHSH violation}) shows not only that
Bell's particular correlations were to be expected, but also what they
depend on. They depend counterfactually but not causally on the quantum
state $\Phi ^{+}$, and they also depend counterfactually on that state's
backing conditions, as described by true magnitude claims. The status of the
quantum state disqualifies it from participation in causal relations, but
true magnitude claims may be taken to describe beables recognized by quantum
mechanics. To decide which conditions backing any of the states involved in
their explanation describe causes of Bell's particular correlations or the
events they correlate we need to return to the connection between causation
and chance.

\qquad The intuition that, other things being equal, a cause raises (or at
least alters) the chance of its effect is best cashed out in terms of an
interventionist counterfactual: $c$ is a cause of $e$ just in case $c,e$ are
distinct actual events and there is some conceivable intervention on $c$
whose occurrence would have altered the chance of $e$. Such an intervention
need not be the act of an agent: it could involve any modification in $c$ of
the right kind. Woodward [2003, p. 98] is one influential attempt to say
what kind of external influence this would involve. Note that Einstein's
formulation of a principle of local action also appeals to intervention:

\begin{quote}
The following idea characterizes the relative independence of objects far
apart in space ($A$ and $B$): external influence on $A$ has no \textit{%
immediate} ("unmittelbar") influence on $B$; this is known as the `principle
of local action' (Einstein [1948, pp. 321-2])
\end{quote}

I used the idea of intervention to argue against any causal dependence
between events in $\mathbf{1}$ and $\mathbf{2}$: anyone who accepts quantum
mechanics accepts that it makes no sense to speak of an intervention in one
of these regions that would affect a hypothetical agent's estimated chance
of what happens in the other. So even though the outcome $e_{B}$ in $\mathbf{%
2}$ backs the assignment $|V_{B}\rangle $ to $L$ at $q$, the outcome\ in $%
\mathbf{1}$ does not depend causally on $e_{B}$: for similar reasons,
neither does the outcome in $\mathbf{2}$\ depend on that in $\mathbf{1}$.
The same idea can now be used to show that both these outcomes \textit{do}
depend causally on whatever event $o$ in the overlap of the backward light
cones of $\mathbf{1}$ and $\mathbf{2}$\ warranted assignment of state $\Phi
^{+}$---an event truly described by magnitude claims that backed this
assignment.

\qquad Assume first that the events $a,b$ at which the polarizers are set on
a particular occasion occur in the overlap of the backward light cones of $%
\mathbf{1}$ and $\mathbf{2}$: this assumption will later be dropped. Let $r$
be a point outside the future light cones of $e_{A},e_{B}$ but within the
future light cone of the event $o$. Let $e_{A}\uplus e_{B}$ be the event of
the joint occurrence of\ $e_{A},e_{B}$. This is an event of a type to which
the Born rule is applicable: the application yields its chance $%
Ch_{r}(e_{A}\uplus e_{B})=$ $Pr_{a,b}^{\Phi ^{+}}(V_{A},V_{B})=%
{\frac12}%
\cos ^{2}\angle ab$. We already saw that $Ch_{r}(e_{A})=\Pr
{}_{_{a,b}{}}^{\Phi ^{+}}(V_{A})=%
{\frac12}%
=\Pr {}_{_{a,b}{}}^{\Phi ^{+}}(V_{B})=Ch_{r}(e_{B})$. The event $o$ affects
all these chances: had a different event $o^{\prime }$ occurred backing the
assignment of a different state (e.g. $|H_{A}\rangle |V_{B}\rangle $), or no
event backing any state assignment, then any or all of these chances could
have been different. Since it makes sense to speak of an agent altering the
chance of event $o$ at $s$ in its past light cone, we have%
\begin{gather*}
Ch_{r}(e_{A}\uplus e_{B}|do-o)\neq Ch_{r}(e_{A}\uplus e_{B}) \\
Ch_{r}(e_{A}|do-o)\neq Ch_{r}(e_{A}) \\
Ch_{r}(e_{B}|do-o)\neq Ch_{r}(e_{B})
\end{gather*}%
where $do-o$ means $o$ is the result of an intervention without which $o$
would not have occurred. It follows that $e_{A},e_{B},e_{A}\uplus e_{B}$ are
each causally dependent on $o$: $o$ is a common cause of $e_{A},e_{B}$ even
though the probabilities of events of these types do not factorize. The same
reasoning applies to each registered photon pair on any occasion at any
settings $a,b$. So the second requirement on explanation is met: the
separate recording events, as well as the event of their joint occurrence,
depend \textit{causally} on the event $o$ that serves to back assignment of
state $\Phi ^{+}$ to the photon pairs involved in this scenario.

\qquad By rejecting any possibility of an intervention expressed by $%
do-e_{B} $ or $do-\bar{e}_{B}$, anyone accepting quantum mechanics should
deny that $Ch_{p(q)}(e_{A}|do-e_{B})\neq Ch_{p(q)}(e_{A}|do-\bar{e}_{B})$ is
true or even meaningful. Nevertheless $Ch_{q}(e_{A}|e_{B})\neq Ch_{q}(e_{A}|%
\bar{e}_{B})$: in this sense $e_{A}$ depends counterfactually but not
causally on $e_{B}$. Does such counterfactual dependence provide reason
enough to conclude that $e_{A}$ is part of the explanation of $e_{B}$? An
obvious objection is that because of the symmetry of the situation with $%
\mathbf{1}$ and $\mathbf{2}$ space-like separated there is an equally strong
reason to conclude that $e_{B}$ is part of the explanation of $e_{A}$,
contrary to the fundamentally asymmetric nature of the explanation relation.
But one can see that this objection is not decisive by paying attention to
the contrasting epistemic perspectives associated with the different
physical situations of hypothetical agents Alice* and Bob* with world-lines
confined to interiors of the light cones of $\mathbf{1}$, $\mathbf{2}$
respectively.

\qquad As his world-line enters the future light cone of $\mathbf{2}$ Bob*
comes into position to know the outcome at $\mathbf{2}$ while still
physically unable to observe the outcome at $\mathbf{1}$. His epistemic
situation is then analogous to that \ of a hypothetical agent Chris in a
world with a Newtonian absolute time, in a position to know the outcome of
past events but physically unable to observe any future event. Many have
been tempted to elevate the epistemic asymmetry of Chris's situation into a
global metaphysical asymmetry in which the future is open while the past is
fixed and settled. It is then a short step to a metaphysical view of
explanation as a productive relation in which the fixed past gives rise to
the (otherwise) open future, either deterministically or stochastically.%
\footnote{%
See, for example, Maudlin [2007, pp. 173-8].}

\qquad Such a move from epistemology to metaphysics should always be treated
with deep suspicion. But in this case it is clearly inappropriate in a
relativistic space-time since the "open futures" of agents like Alice* and
Bob* cannot be unified into \textit{the} open future. This prompts a retreat
to a metaphysically drained view of explanation as rooted in cognitive
concerns of a physically situated agent, motivated by the need to unify,
extend and efficiently deploy the limited information to which it has access.

\qquad For many purposes it is appropriate to regard the entire scientific
community as a (spatially) distributed agent, and to think of the provision
of scientific explanations as aiding \textit{our} collective epistemic and
practical goals. This is appropriate insofar as localized agents share an
epistemic perspective, with access to the same information about what has
happened. But Alice* and Bob* do not have access to the same information at
time $t_{2}$ or $t_{3}$ since they are then space-like separated. So it is
entirely appropriate for Bob* to use $e_{B}$ to explain $e_{A}$ and for
Alice* to use $e_{A}$ to explain $e_{B}$. This does not make explanation a
subjective matter for two reasons. There is an objective physical difference
between the situations of Alice* and Bob* underlying the asymmetry of their
epistemic perspectives: and by adopting either perspective in thought (as I
have encouraged the reader to do) anyone can come to appreciate how each
explanation can help make Bell's correlations seem less puzzling.
Admittedly, neither explanation is very deep, and I will end by noting one
puzzle that remains.

\qquad By meeting both minimal requirements on explanation, the application
of quantum theory enables us to explain Bell's correlations. But 
\QTR{frametitle}{is} this explanation local? Several senses of locality are
relevant here. \QTR{frametitle}{T}he explanation involves no superluminal
causal dependence. As stated, the condition of Local Causality is not
applicable to the quantum mechanical explanation since it presupposes the
uniqueness of the probability to which it refers. (\ref{FactorizabilitySU})
(and presumably also (\ref{Factorizability})) are violated, but Bell ([2004,
p. 243]) preferred to see (\ref{Factorizability}) as not a formulation but a
consequence of `local causality': I have argued that it is not. To retain
its connection to (IP), a version of Local \ Causality should speak of
chances rather than general probabilities. A version that equates the
unconditional chance of $e_{A}$ to its chance conditional on $e_{B}$ holds,
no matter how these chances are relativized to the same space-time point.
But a version that is clearly motivated by the intuitive principle (IP)
would rather equate the unconditional chance of $e_{A}$ to its chance
conditional on \textit{an intervention that produces} $e_{B}$. However this
version is inapplicable since acceptance of quantum mechanics renders
senseless talk of interventions producing $e_{B}$.

\qquad The explanation one can give by applying quantum mechanics appeals to 
\textit{chances} that are localized, insofar as they are assigned at
space-time points that may be thought to offer the momentary perspective of
a hypothetical idealized agent whose credences they would guide. But these
chances are not quantum beables and they are not physical propensities
capable of manifestation at those locations (or anywhere else). The only
causes figuring in the explanation are localized where the physical systems
are whose magnitudes back the assignment of state $\Phi ^{+}$. That chances
are not propensities becomes clear when one drops the assumption that $a,b$
occur in the overlap of the backward light cones of $\mathbf{1}$ and $%
\mathbf{2}$, as depicted in Figure 5. \FRAME{fhF}{3.4549in}{3.039in}{0pt}{}{%
}{Figure}{\special{language "Scientific Word";type
"GRAPHIC";maintain-aspect-ratio TRUE;display "USEDEF";valid_file "T";width
3.4549in;height 3.039in;depth 0pt;original-width 5.8816in;original-height
5.1673in;cropleft "0";croptop "1";cropright "1";cropbottom "0";tempfilename
'NDR1RI09.wmf';tempfile-properties "XPR";}}

If $a,b$ are set at the last moment, the chance of $e_{A}\uplus e_{B}$ that
figures in its explanation may be located \textit{later} in the laboratory
frame than $e_{A},e_{B}$. If chance were a physical propensity it should act 
\textit{before} its manifestation. But chances aren't\textit{\ }%
propensities---proximate causes of localized events. They are a localized
agent's objective guide to credence about epistemically inaccessible events.

\qquad I will conclude by noting one sense in which the explanation one\ can
give using quantum mechanics is not local as it stands. Though it is a
(non-factorizable) cause of \ events of types $A,B$ in regions $\mathbf{1,}$ 
$\mathbf{2}$ respectively, the event $o$ is not connected to its effects by
any spatiotemporally continuous causal process described by quantum
mechanics. This puts the explanation in tension with the\textit{\ first }%
conjunct of Bell's ([2004, p. 239]) intuitive locality principle (IP):%
\newline
\textit{"The direct causes (and effects) of events are near by}, and even
the indirect causes (and effects) are no further away than permitted by the
velocity of light."

$o$ is separated from both recording events in regions $\mathbf{1,}$ $%
\mathbf{2}$ in time, and from at least one in space. If $o$ is not merely a
cause but a \textit{direct} cause of these events then it violates the first
conjunct of (IP) because it is \textit{not} nearby. But if one adopts the
present view of quantum mechanics, the theory has no resources to describe
any causes mediating between $o$ and these recording events. So while their
quantum explanation is not explicitly inconsistent with the first conjunct
of (IP), mediating causes could be found only by constructing a new theory.
Bell's work has clearly delineated the obstacles that would have to be
overcome on that path.

\bigskip

\end{document}